\documentstyle[preprint,epsf,aps]{revtex}
\newcommand{\Ne}{N\`{e}el}
\newcommand{\be}{\begin{equation}}
\newcommand{\ee}{\end{equation}}
\begin{document}
\title{Magnetic model for $Ba_2Cu_3O_4Cl_2$}

\author{J.Richter, A.Voigt, J.Schulenburg, N.B.Ivanov$^*$ and R.
Hayn$^{**}$}

\address{Institut f\"ur Theoretische Physik,
Otto-von-Guericke-Universit\"at Magdeburg, Postfach 4120, 39106
Magdeburg, Germany\\ 
$^*$ Georgi Nadjakov Institute for Solid State Physics, Bulgarian
Academy of Sciences, 72 Tzarigradsko chaussee blvd., 1784 Sofia,
Bulgaria \\
$^{**}$ Fachbereich Physik, TU Dresden, Germany}

\date{\today}
\maketitle

\begin{abstract}
$Ba_2Cu_3O_4Cl_2$ consists of two types of copper atoms, $Cu(A)$ and
$Cu(B)$. We study the corresponding Heisenberg model with three
antiferromagnetic couplings, $J_{AA}$, $J_{BB}$ and $J_{AB}$. We find
interesting frustration effects due to the coupling $J_{AB}$.
\end{abstract}

Keywords: antiferromagnetism, quantum fluctuations, frustration,
phase transition


\section{Introduction}
The exciting collective magnetic properties of layered cuprates have
attracted much attention over the last decade. Recent experiments on
$Ba_2Cu_3O_4Cl_2$ \cite{ref1} show magnetic properties of this quasi 2d
quantum antiferromagnet which differ from the well studied
antiferromagnetism of the parent cuprates like  $La_2CuO_4$. Most
interesting is the observation of two magnetic critical temperatures
($T^c_A \sim 330K$, $T^c_B \sim 40K$) and of a weak ferromagnetic moment
\cite{ref1}, where a simple explanation of the ferromagnetic moment by a
Dzyaloshinsky-Moriya exchange can be ruled out \cite{ref2}.

The important difference between the parent cuprates and
$Ba_2Cu_3O_4Cl_2$ is the existence of additional $Cu(B)$ atoms located
at the centre of every second $Cu(A)$ square. The coupling between
the $A$ spins $J_{AA}$ is strongly antiferromagnetic. The observation of
a smaller second critical temperature $T^c_B$ \cite{ref1} indicates a
weaker antiferromagnetic coupling $J_{BB}$ between the $B$ spins.
Additionally, there is a competing coupling $J_{AB}$ between $A$ and $B$
spins giving rise for interesting frustration effects.

              
\section{The Model}

We consider the classical and the quantum (spin $1/2$) version of the 
Heisenberg model 

\be
H=  
J_{AA} \hspace{-12pt} \sum_{<i \in A, j \in A>}\hspace{-12pt}{\bf S_iS_j}
\hspace{4pt}+\hspace{4pt}
J_{BB} \hspace{-12pt}\sum_{<i \in B, j \in B>}\hspace{-12pt}{\bf S_iS_j}
\hspace{4pt}+\hspace{4pt}
J_{AB} \hspace{-12pt}\sum_{<i \in A, j \in B>}\hspace{-12pt}{\bf S_iS_j}
\label{ham}
\ee

where the sums run over nearest neighbour bonds of type $A-A$, $B-B$,
$A-B$. Tight binding calculations \cite{ref3} and the large $T^c_A$
indicate a strong antiferromagnetic $J_{AA}$. Because the couplings
$J_{BB}$ and $J_{AB}$ are of higher order of the hopping integrals
\cite{ref3} they can be assumed as (much) smaller then $J_{AA}$. In this
paper we focus our interest on  antiferromagnetic $J_{BB}$, $J_{AB} <
J_{AA}$ and discuss the magnetic properties in dependence on $J_{AB}$
and $J_{BB}$  for fixed $J_{AA}=1$. Since the copper spin is $1/2$ the
quantum nature of the spins is of great importance. To take into account
full quantum fluctuations we use an exact diagonalization algorithm to
calculate data for two finite lattices of $N=12$ sites ($8$ $A$ spins
and $4$ $B$ spins) $N=24$ sites ($16$ $A$ spins and $8$ $B$ spins). For
comparison we present also some data for the corresponding classical
systems.

\section{Results}

The ground state results for small $J_{BB}$ and $J_{AB}$ are as follows:

{\it - Classical model:}
Starting from the point $J_{AB}=0$ we have \Ne\ ordering in the
subsystems A and B. These two \Ne\ states are decoupled and can rotate
freely with respect to each other, i.e. the ground state is highly
degenerated.

Increasing the frustrating $J_{AB}$ there is a first order transition to
a non-planar canted state at $J^c_{AB}=2\sqrt{J_{AA}J_{BB}}$ (see
Fig.1). In this non-planar state the $A$ spins form a slightly tilted
\Ne\ state, where  the sublattice magnetization $M_{s,A}$ is lowered  to
about 90\% for $J_{AB}\stackrel{>}{\sim} J^c_{AB}$ and decreases with
increasing $J_{AB}$. In the $B$ subsystem the angle between neighbouring
spins is $\pi / 2$ and the spins built a planar state perpendicular to
the sublattice magnetization axis of the $A$ subsystem. However, due to
the canting of $A$ spins there is a correlation between $A$ and $B$
spins. There is no ferromagnetic moment in the non-planar canted state.
However, an almost degenerated classical planar state  exists with a
weak ferromagnetic moment which increases with $J_{AB}$. Quantum
fluctuations may change this situation and could select a planar state
instead of a non-planar ground state. For larger $J_{AB}$ further
transitions to other ground states take place, which are not considered
here.

{\it - Quantum model:} 
We focus our consideration on the lattice with $N=24$ sites. The results
for $N=12$ are comparable supposing a relation of $J^{12}_{BB}=2
J^{24}_{BB}$. (Notice, that for $N=12$ the $B$-spins have only two
nearest $B$ neighbors instead of four.) First we consider the
antiferromagnetic ground state ordering  for small $J_{AB}$, where the
two subsystems order in a quantum \Ne\ state. In contrast to the
classical case the quantum fluctuations cause a typical 'order from
disorder' effect and lift the classical degeneracy by selecting a state
with a collinear structure of two sublattice magnetizations. This is
accompanied by the development of a magnetic coupling between $A$ and
$B$ subsystems. A similar 'order from disorder' phenomenon is well-known
from the $J_1 - J_2$ antiferromagnet on square lattice for $J_2/J_1 \sim
0.65$ \cite{ref4}. The stability  of the \Ne\ state is supported by
quantum fluctuations (Fig.1). The transition line to a canted state lies
well above the classical instability line and for small $J_{BB} < 0.07
J_{AA}$ the line follows approximately the relation $J^c_{AB} \approx
2.7 \sqrt{J_{AA}J_{BB}}$. A detailed analysis of the canted state gives
indications for planar structure which is, however, slightly different
from that classical planar state which is almost degenerated with the
non-planar classical ground state. The ground state order parameters for
the $N=24$ site lattice are shown in Fig.2 for $J_{BB}=0.1J_{AA}$ which
corresponds to the relation of critical temperatures. Obviously, both
systems are antiferromagnetically ordered. However, the
antiferromagnetic order in the $B$ system is weakened by $J_{AB}$ and
drops down dramatically at the critical line.  This is accompanied by an
increase of the square of the total magnetic moments $M^2_B$, $M^2_A$
indicating the possibility of a weak ferromagnetic moment in the quantum
system.

Let us finally present the temperature dependence of the specific heat
$c(T)$ for the quantum case (Fig.3). The exact calculation of $c(T)$
needs the complete diagonalization of the Hamiltonian and is restricted
to very small system, i.e. to $N=12$ in our case. We find two peaks in
$c(T)$ indicating the two transition temperatures. The peak positions
$T_A$ and $T_B$ correspond to the coupling strengths $J_{AA}$ and
$J_{BB}$. The frustrating coupling $J_{AB}$ causes a decrease of $T_A$
and $T_B$. For the parameters used in Fig.3 we have
$T_A(J_{AB}=0.5)/T_A(J_{AB}=0) = 0.99$ and
$T_B(J_{AB}=0.5)/T_B(J_{AB}=0) = 0.87$.


\section{Acknowledgments}
This work has been supported by the Deutsche Forschungsgemeinschaft
(Project No. Ri 615/1-2) and the Bulgarian Science Foundation (Grant
F412/94).



\begin{figure}
\caption{
Transition line between the \Ne\ phase and the canted phase (see text)
for small $J_{AB}$ and $J_{BB}$ ($N=24$). Notice that for larger
$J_{AB}$ several other phases appear which are not considered here.
\label{fig1}
}
\end{figure}

\begin{figure}
\caption{
Square of magnetic order parameters versus $J_{AB}$ for $J_{BB}=0.1$
($N=24$). $M^2_{s,A}$ ($M^2_{s,B}$) - staggered magnetic moment of
subsystem $A$ ($B$), $M^2_{A}$ ($M^2_{B}$) - total magnetic moment  of
subsystem $A$ ($B$)
\label{fig2}
}
\end{figure}

\begin{figure}
\caption{
Specific heat versus temperature for $J_{BB}=0.25$ ($N=12$) and two
different  $J_{AB}$.
\label{fig3}
}
\end{figure}


\begin{thebibliography}{99}
\bibitem{ref1}
S.Noro et al., Mater.Sci.Eng. B {\bf 25} (1994), 167;
K.Yamada et al. Physica B {\bf 213-214} (1995), 191.

\bibitem{ref2} F.C.Chou et al. Phys.Rev.Let. {\bf 78} (1997), 535.
 
\bibitem{ref3}
H.Rosner and R.Hayn,  Physica B (1997), to be published.

\bibitem{ref4} K.Kubo, T.Kishi, J.Phys.Soc.Jap.
  {\bf 60} (1991), 567;
 J.Richter, Phys.Rev.B {\bf 47} (1993), 5794.

\end{thebibliography}
\end{document}